\documentclass{article}
\usepackage{amssymb}
\usepackage{amsmath}
\usepackage{hyperref}
\usepackage{amsfonts}
\usepackage{graphicx}
\usepackage{authblk}

\begin{document}

\title{Quantum to classical limit of open systems}

\author[1]{G. Bellomo}
\author[2]{M. Castagnino}
\author[3]{S. Fortin\thanks{sfortin@gmx.net}}
\affil[1]{\small IAFE (CONICET-UBA) and FCEN (UBA), Casilla de Correos 67, Sucursal 28, 1428, Buenos Aires, Argentina.}
\affil[2]{\small CONICET, IAFE (CONICET-UBA), IFIR and FCEN (UBA), Casilla de Correos 67, Sucursal 28, 1428, Buenos Aires, Argentina.}
\affil[3]{\small CONICET, IAFE (CONICET-UBA) and FCEN (UBA), Casilla de Correos 67, Sucursal 28, 1428, Buenos Aires, Argentina.}

\date{\today }
\maketitle

\begin{abstract}
We present a complete review of the quantum-to-classical limit of open systems by means of the theory of decoherence and the use of the Weyl-Wigner-Moyal (WWM) transformation. We show that the analytical extension of the Hamiltonian provides a set of poles that can be used to (a) explain the non-unitary evolution of the relevant system and (b) completely define the set of preferred states that constitute the mixture into which the system decoheres: the Moving Preferred Basis. Moreover, we show that the WWM of these states are the best candidates to obtain the trajectories in the classical phase-space.
\end{abstract}

\section{Introduction}

The quantum-to-classical transition has been exhaustively studied from the very conception of the quantum theory. At first sight, the intrinsic probabilistic structure and the nonlocal correlations of the quantum reality seem incompatible with the classical world. So one of the main issues was the acceptance of the quantum theory as universal. In this sense, the development of the theory of decoherence, mainly the Environment-Induced Decoherence (EID) approach, provided a new and insightful way of thinking about this limit \cite{Zur(02), Zur(03), Zur(07), Sch(07), Sch(08), Joo(85), Joo(03), Omn(94)}.

Now it is well established that the transition is closely interrelated to the problem of the nonobservability of interference. To account for the emergence of classicality from quantum mechanics it is necessary to introduce a mechanism that `eliminates' the quantum coherence, at least for the case of macroscopic systems. In addition to providing this mechanism for open systems, decoherence introduces the notion of preferred states \cite{Zur(81)} -the ones that get less entangled with the environment. This states form the basis set into which the quantum system \emph{actualize}, thus representing the best candidates to explain (at least, the appearence of) the classical reality. This characterization of the preferred states as the least entangled with the environment is the key of the operational \emph{predictability sieve criterion} for picking the pointer basis.

In this paper, we show that the description of the (non-unitary) evolution of an open system in terms of the poles' catalogue associated to the analytical extension of the Hamiltonian provides a well defined method to obtain the set of preferred states. First, in section \ref{sec:decoherence} we make a brief overview of the main concepts related to the decoherence of open systems and the `problem' of the preferred basis. Section \ref{sec:quantum_phasespace} discusses the mapping of the quantum space of operators into a phase-space by the Weyl-Wigner-Moyal symbol. In section \ref{sec:classical_limit} we show in detail that these concepts provide a new approach for obtaining the classical trajectories. Summary and conclusions are presented in section \ref{sec:conclusions}.

\section{Decoherence and the Preferred Basis}
\label{sec:decoherence}

Decoherence is mostly presented as the mechanism by which the coherence of the state of a quantum open system entangled with its environment is lost. This approach considers the partition of the whole into a proper (or relevant)system ${\mathcal{S}}$ and its environment ${\mathcal{E}}$. Into this formalism, one defines the \emph{reduced density operator} $\widehat{\rho}_{\mathcal{S}}={\text{tr}}_{\mathcal{E}}{\text{\thinspace}}\widehat{\rho}$, where $\widehat{\rho}$ is the density operator of the whole system and \ `tr$_{\mathcal{E}}$' refers to the \ `partial trace' over the environmental degrees of freedom. Even when $\widehat{\rho}$ evolves unitarily, obeying the Liouville-von Neumann equation, the reduced density operator evolves in a non-unitary manner.

Usually, the irreversible character of the evolution of $\widehat{\rho}_{\mathcal{S}}$ is related to two different processes. First, one encounters that there exists a set of states of ${\mathcal{S}}$, namely the \emph{preferred states}, that defines a basis $\{\left\vert {j(t)}\right\rangle \}$ in which $\widehat{\rho}_{\mathcal{S}}$ becomes \emph{diagonal} after a characteristic time of decoherence $t_{D}$. Second, the evolution of the proper system reaches a stationary period after a characteristic time of relaxation $t_{R}\gg t_{D}$. Thus, we define
\begin{description}
	\item[Decoherence] as the non-unitary evolution of $\widehat{\rho}_{\mathcal{S}}$ toward a state that is diagonal in some well defined \emph{preferred basis} $\{\left\vert {j(t)}\right\rangle \}$, for $t>t_{D}$.
	\item[Relaxation] as the non-unitary evolution of $\widehat{\rho}_{\mathcal{S}}$ towards an equilibrium state $\widehat{\rho}_{\ast}$ at a typical \emph{relaxation time} $t_{R}$, with $\widehat{\rho}_{\ast}$ diagonal in its own eigen-basis $\{\left\vert {i_{\ast}}\right\rangle \}$ (therelaxation basis).
\end{description}

Schematically, the proper system evolves as $\widehat{\rho}_{\mathcal{S}}(t)\longrightarrow\widehat{\rho}_{PS}(t)\longrightarrow\widehat{\rho}_{\mathcal{S\ast}}$, where $\widehat{\rho}_{PS}(t)$ that corresponds to the decohered system will be called the \emph{privileged state}.

The preferred states are shown to be the best candidates to constitute the set of `quasi-classical states', in the sense that its correlations are less affected by the entanglement with the environment. This set is not rigorously defined in the literature, but is commonly characterized by the properties of \emph{robustness} and \emph{minimization of the production of entropy}, in the senses that will be explained in section \ref{sec:mpb_criteria}. In most examples, these preferred states turn out to be representative of some collective variables of the relevant system.

An open system ${\mathcal{S}}$ can always be considered as part of a (larger) closed system ${\mathcal{U}}$ decomposed as ${\mathcal{U}}={\mathcal{S}}\otimes{\mathcal{E}}$. In order to obtain the non-unitary evolution in the EID approach, one must consider some \emph{relevant observables} of the type $\widehat{O}_{R}=\widehat{O}_{\mathcal{S}}\otimes\widehat{I}_{\mathcal{E}}$, where $\widehat{O}_{\mathcal{S}}$ is any observable of the relevant (or proper) system ${\mathcal{S}}$ and $\widehat{I}_{\mathcal{E}}$ is the unit operator in the Hilbert space of ${\mathcal{E}}$. It can be proved in a straightforward manner that
\begin{equation}
	\left\langle \widehat{O}_{R}\right\rangle _{\rho(t)}=\text{tr}(\widehat{\rho}(t)\widehat{O}_{R})=\text{tr}(\widehat{\rho}_{\mathcal{S}}(t)\widehat{O}_{\mathcal{S}})=\left\langle\widehat{O}_{\mathcal{S}}\right\rangle_{\rho_{\mathcal{S}}(t)}
	\label{eq:eid_partial_trace}
\end{equation}
where $\widehat{\rho}_{\mathcal{S}}(t)={\text{tr}}_{\mathcal{E}}\widehat{\rho}(t)$, so the last member of the last equation is computed in the Hilbert space of ${\mathcal{S}}$ to give the expectation value of the observable $\widehat{O}_{R}$ for the system in the state $\widehat{\rho}$.

\subsection{The Poles Technic and the MPB}
\label{sec:poles_technic}

To introduce our generall definition of the preferred basis we will follow the reasoning of Refs. \cite{Cas(10), Cas(11)}. We start by stating that, given a system, all the possible decaying modes related to its non-unitary evolution are contained in the poles' catalogue obtained through the analytical continuation of its Hamiltonian. This is essential to understand that relaxation and decoherence must arise from this catalogue. So, to obtain the typical non-unitary modes of the evolution of $\widehat{\rho}_{\mathcal{S}}$ it is usual to extend the range of the eigenvalues of the Hamiltonian from the real semiaxis to the complex plane, obtaining the complex eigenvalues
\begin{equation}
	z_{i}=\omega_{i}-\frac{i}{2}\gamma_{i}
	\label{eq:complex_ev}
\end{equation}
with $\omega_{i}$ the real energy eigenvalues and $\gamma_{i}$ the frequencies corresponding to the decaying modes. Also, $z_{i}$ are the complex poles of the resolvent or those of the complex extension of the S-matrix (see, e.g., Refs. \cite{Mos(09), Lon(09)}). Each mode corresponds to a characteristic decaying time $\tau_{i}=\frac{\hbar}{\gamma_{i}}$ and there is also a `long time' (or Khalfin) decaying mode \cite{Rot(06)}. Usually, this last mode can be neglected for all practical purposes because of its extreme length. In that case, it can be proved that \cite{Cas(01), Cas(02)}
\begin{equation}
	\left\langle {\widehat{O}_{\mathcal{S}}}\right\rangle =\text{tr}(\widehat{\rho}_{\mathcal{S}}(t)\widehat{O}_{\mathcal{S}})=\text{tr}({\widehat{\rho}_{{\mathcal{S}}\ast}\widehat{O}_{\mathcal{S}}})+\sum_{i=0}^{N}{a_{i}(t)\operatorname{exp}\left(  -\frac{\gamma_{i}}{\hbar}t\right)}
	\label{eq:evol_modos}
\end{equation}
where $\widehat{\rho}_{{\mathcal{S}}\ast}$ is the equilibrium state of the proper system and the $a_{i}(t)$ are real oscillating coefficients that can be computed from the data of the system and the initial conditions. If $\gamma_{0}={\text{Im}}(z_{0})$ is the minimum of the $\left\{\gamma_{i}\right\}$, it is quite obvious that $z_{0}$ is the pole closest to the real axis and therefore it defines the relaxation time
\begin{equation}
	t_{R}=\frac{\hbar}{\gamma_{0}}.
	\label{eq:relaxation_time}
\end{equation}

In order to introduce the MPB, lets consider the general case with $N$ poles, i.e. $N$ decaying modes determined by $\gamma_{0}<\gamma_{1}<...<\gamma_{N}$. Then, the decoherence time is defined as \cite{Cas(10)}
\begin{equation}
	t_{D}=\frac{\hbar}{\gamma_{eff}}
	\label{eq:decoherence_time}
\end{equation}
where
\begin{equation}
	\gamma_{eff}=\frac{\sum_{i=0}^{N}{a_{i}(0)\gamma_{i}}}{\sum_{i=0}^{N}{a_{i}(0)}}
	\label{eq:effective_gamma}
\end{equation}
is defined as an \emph{effective mode} such that the modes corresponding to $\gamma_{i}<\gamma_{eff}$ (i entre 1 y M) are called the \emph{slow modes} while the ones with $\gamma_{i}<\gamma_{eff}$ (i entre M+1 y N) are called the \emph{fast modes}. Eq. (\ref{eq:evol_modos}) suggests the definition of a \emph{privileged state,} $\widehat{\rho}_{PS}(t)$, such that
\begin{equation}
	 \text{tr}({\widehat{\rho}_{PS}(t)\widehat{O}_{\mathcal{S}}})=\text{tr}({\widehat{\rho}_{{\mathcal{S}}\ast}\widehat{O}_{\mathcal{S}}})+\sum_{i=0}^{M}{a_{i}(t)\operatorname{exp}\left(-\frac{\gamma_{i}}{\hbar}t\right)}
	\label{eq:preferred_state_functional}
\end{equation}
where the sum runs over the $M<N$ slow modes. If we choose an exhaustive set of observables, Eq.(\ref{eq:preferred_state_functional}) completely defines $\widehat{\rho}_{PS}(t)$
\begin{equation}
	\widehat{\rho}_{PS}(t)=\widehat{\rho}_{{\mathcal{S}}\ast}+\sum_{i=0}^{M}\widehat{a}_{i}(t)e^{-\frac{\gamma_{i}}{\hbar}t}
	\label{eq:PS_entropy}
\end{equation}
with $\widehat{a}_{i}(t)$ dependent on the chosen observables. The eigen-decomposition of $\widehat{\rho}_{PS}(t)$
\begin{equation}
	\widehat{\rho}_{PS}(t)=\sum\nolimits_{i}p_{i}(t)\left\vert {i_{PS}(t)}\right\rangle \left\langle {i_{PS}(t)}\right\vert.				 \label{eq:preferred_state}
\end{equation}
defines a time-dependent basis $\left\{  \left\vert {i_{PS}(t)}\right\rangle\right\}  $ which diagonalizes $\widehat{\rho}_{PS}(t)$, the so called MPB. The time dependence of the $p_{i}(t)$'s can be justified appealing at the fact that while the projectors $\left\vert{i_{PS}(t)}\right\rangle \langle{i_{PS}(t)|}$ must evolve unitarily, the state $\widehat{\rho}_{\mathcal{S}}$ evolves non-unitarily. For $t>t_{D}$ all the fast modes of the unitary evolution of $\widehat{\rho}_{\mathcal{S}}$ becomes negligible and the MPB converges to the eingenbasis of the proper system $\widehat{\rho}_{\mathcal{S}}(t)$, i.e. $\widehat{\rho}_{\mathcal{S}}(t)\approx\widehat{\rho}_{PS}(t)$ for $t>t_{D}$ and therefore $\widehat{\rho}_{\mathcal{S}}$ becomes (almost) diagonal in the MPB.

Let us consider, as mode of example, the Omn\`{e}s (or Lee-Friederich) model. Its Hamiltonian is \cite{Omn(94)}
\begin{equation}
	 \widehat{H}=\omega\widehat{a}{^{\dagger}}\widehat{a}+\int{\omega_{k}\widehat{b}_{k}{^{\dagger}}\widehat{b}_{k}}dk+\int{\lambda_{k}\widehat{a}_{k}{^{\dagger}}\widehat{b}_{k}+\lambda_{k}^{\ast}\widehat{b}_{k}{^{\dagger}}\widehat{a}_{k}}dk
	\label{eq:lf_hamiltonian}%
\end{equation}
where we take an oscillator of frecuency $\omega$ described by its creation an annihilation operators $\widehat{a}$ and $\widehat{a}^{\dagger}$ (the proper system) and a collection of oscillators with creation and annihilation operators $\widehat{b}_{k}$ and $\widehat{b}_{k}^{\dagger}$ and frecuencies $\omega_{k}$ (the environment). The last term in Eq.(\ref{eq:lf_hamiltonian}) represents the coupling between both subsystems, $\lambda_{k}$ being the coupling constants. This Hamiltonian is such that if $\left\vert {\nu}\right\rangle $ is a $\nu$-mode state so it is $\left\vert {\nu^{\prime}}\right\rangle=\widehat{H}\left\vert {\nu}\right\rangle $. Then, the number of `mode sectors' is conserved and it is possible to reduce the problem to the one with only one mode sector. Doing this we can find the pole $z_{0}$ closest to the real axis which, up to second order in $\lambda_{k}$, is \cite{Lau(99)}
\begin{equation}
	z_{0}=\omega+\int{\frac{\lambda_{k}^{2}dk}{\omega-\omega_{k}+i0}}
	\label{eq:lf_pole0}
\end{equation}
where
\begin{equation}
	z_{0}=\omega_{0}^{\prime}-\frac{i}{2}\gamma_{0}
	\label{eq:lf_pole0bis}
\end{equation}
with
\begin{equation}
	\gamma_{0}=\pi\int{n(\omega^{\prime})|\lambda_{\omega^{\prime}}|}^{2}{\delta(\omega-\omega^{\prime})d\omega^{\prime}}
	\label{eq:lf_gamma0}
\end{equation}
and where $dk=n(\omega^{\prime})d\omega^{\prime}$. So, by virtue of Eq.(\ref{eq:relaxation_time}) the relaxation time of this one mode sector is $t_{R}=\frac{\hbar}{\gamma_{0}}$, in exact coincidence to the result obtained by Omn\'{e}s (ref). Now, in the many sector case it is straightforwardly proven that all the poles of the system are $z_{n}=nz_{0}$ with $n=1,2,3,...$, so $z_{0}$ is still the pole closest to the real axis and the relaxation time is given by $t_{R}=\frac{\hbar}{\gamma_{0}}$. Moreover, this suggests the definition of an \emph{effective Hamiltonian} $\widehat{H}_{eff}$ that produces the non-unitary evolution the proper system
\begin{equation}
	\widehat{H}_{eff}=z_{0}\widehat{a}_{0}^{\dagger}\widehat{a}_{0}=z_{0}\widehat{N}
	\label{eq:lf_effective_h}
\end{equation}
where $\widehat{a}_{0}$ and $\widehat{a}_{0}^{\dagger}$ are the annihilation and creation operators of the mode corresponding to the pole $z_{0}$, and $\widehat{N}$ is the operator associated to the number of poles $n$, with $\widehat{N}\left\vert{n}\right\rangle =n\left\vert {n}\right\rangle $. So $\{\left\vert {n}\right\rangle \}$ is the common eigenbasis of $\widehat{N}$ and $\widehat{H}_{eff}$. To compute the decoherence time we need to introduce some intitial conditions. First, consider two coherent states $\left\vert{\alpha_{1}(0)}\right\rangle$ and $\left\vert {\alpha_{2}(0)}\right\rangle$ defined by $\alpha_{i}(0)=\frac{m\omega_{0}^{\prime}}{\sqrt{2m\hbar^{2}\omega_{0}^{\prime}}}x_{i}(0)$, i=1,2. Let $\widehat{\rho}_{\mathcal{S}}(0)=\left\vert {\Phi(0)}\right\rangle \left\langle {\Phi(0)}\right\vert $ with $\left\vert{\Phi(0)}\right\rangle =a\left\vert {\alpha_{1}(0)}\right\rangle +b\left\vert {\alpha_{2}(0)}\right\rangle$ be the initial condition for the reduced density operator. Then, the non-diagonal part of $\widehat{\rho}_{\mathcal{S}}$ is
\begin{equation}
	\widehat{\rho}_{\mathcal{S}}^{ND}(t)=ab^{\ast}\left\vert {\alpha_{1}(t)}\right\rangle \left\langle {\alpha_{2}(t)}\right\vert+a^{\ast}b\left\vert{\alpha_{2}(t)}\right\rangle \left\langle {\alpha_{1}(t)}\right\vert .
	\label{eq:lf_rho_nd}
\end{equation}

Computing the time evolution of $\widehat{\rho}_{\mathcal{S}}^{ND}(t)$ it can be proved that for sufficiently short time \cite{Cas(10)}
\begin{equation}
	\widehat{\rho}_{\mathcal{S}}^{ND}(t)\propto\exp\left(  {-\frac{m\omega_{0}^{\prime}}{2\hbar^{2}}\gamma_{0}L^{2}t}\right)
	\label{eq:lf_rho_nd_2}
\end{equation}
where $L=|x_{2}(0)-x_{1}(0)|$ measures the distance between the initial positions of the coherent states. Finally, from our definition of decoherence time (see Eq. (\ref{eq:decoherence_time})) we obtain
\begin{equation}
	t_{D}=\frac{2\hbar^{2}}{m\omega_{0}^{\prime}\gamma_{0}L^{2}},
	\label{eq:lf_td}
\end{equation}
which in comparison with $t_{R}$ gives $t_{D}=\frac{2\hbar^{2}}{m\omega_{0}L^{2}}t_{R}$, again in coincidence with Omn\`{e}s result \cite{Omn(94)}.

\subsection{Entropy production and robustness}
\label{sec:mpb_criteria}

Usually, the MPB is defined as the set of states that are the \emph{most robust} under the influence of the environment. Alternatively, its characterized as the choice that \emph{minimizes the linear entropy production}. Both approaches relies in the fact that the preferred states are those less entangled with the environment. Our definition of the MPB\ satisfies both conditions.

Given a state $\widehat{\rho}$, its \emph{linear entropy} is defined as \cite{Joo(03)}
\begin{equation}
	S_{lin}(t)={\text{tr\thinspace}}(\widehat{\rho}-\widehat{\rho}^{2})
	\label{eq:linear_entropy}
\end{equation}
and the variation of entropy
\begin{equation}
	\frac{d}{dt}S_{lin}(t)=-\frac{d}{dt}{\text{tr\thinspace}}\widehat{\rho}^{2}(t)=-2{\text{tr\thinspace}}(\widehat{\rho}\dot{\widehat{\rho}})
	\label{eq:entropy_variation}
\end{equation}
is shown to be a measurement of the robustness. The task is to find the states that minimizes Eq. (\ref{eq:entropy_variation}) in order to identify the candidates for the preferred basis. In our formalism, this is trivially accomplished since the preferred states are \emph{pure states}: the set of projectors $\{\widehat{\Pi}_{i}\}$ with $\widehat{\Pi}_{i}=\left\vert{i_{PS}(t)}\right\rangle\langle{i_{PS}(t)|}$. Moreover the system evolves towards a privileged state defined in terms of the slow modes only (see Eq.(\ref{eq:PS_entropy})). In consequence, $\frac{d}{dt}S_{lin}(t)$ would be minimal for the entropy of the reduced state.

Otherwise, consider a Hamiltonian $\widehat{H}_{0}$ with no poles, such that in the complete Hamiltonian $\widehat{H}=\widehat{H}_{0}+\widehat{V}$ it is the interaction $\widehat{V}$ the one that produces the poles. Which would be the decaying modes `less affected by the interaction'? Obviously, the ones corresponding to the poles closer to the real semiaxis: the slow modes that produces the slowest decaying factors $e^{-\frac{\gamma_{i}}{\hbar}t}$.

In this way, our pole method to define the MPB is in no contradiction with the traditional literature on the subject if the notions of minimization of the linear entropy and robustness are defined as below.

\section{WWM mapping and the phase-space}
\label{sec:quantum_phasespace}

Let $\mathcal{M}\equiv{\mathbb{R}}^{2N}$ be the phase space of our classical system. Then, the algebra $\widehat{{\mathcal{A}}}$ of regular operators $\widehat{O}$ of the quantum system can be mapped on ${\mathcal{A}}_{q}$, the algebra of ${\mathbb{L}}_{1}$ functions over $\mathcal{M}$, via the \emph{Weyl-Wigner-Moyal symbol}
\begin{equation}
	{\text{symb}}\,:\,\widehat{{\mathcal{A}}}\rightarrow{\mathcal{A}}_{q},\hspace{1cm}{\text{symb}}\,\widehat{O}=O(\phi)
	\label{eq:symbol}
\end{equation}
where $\phi$ symbolizes the coordinates over $\mathcal{M}$.

Given an operator $\widehat{f}\in\widehat{{\mathcal{A}}}$ its associated function $f\in{\mathcal{A}}_{q}$ is obtained trought the \emph{Wigner transformation}
\begin{equation}
	{\text{symb}}\,\widehat{f}\circeq f(\phi)=\int\langle{q+\Delta}|{\widehat{f}}|{q-\Delta}\rangle e^{2i\frac{p\Delta}{\hbar}}d\Delta^{N}
	\label{eq:wigner}
\end{equation}
The operation on ${\mathcal{A}}_{q}$ related with the multiplication on $\widehat{{\mathcal{A}}}$ is the \emph{star product}, defined as
\begin{equation}
	{\text{symb}}\,(\widehat{f}\widehat{g})={\text{symb}}\,\widehat{f}\star{\text{symb}}\,\widehat{g}=(f\star g)(\phi).
	\label{eq:symb_product}
\end{equation}
Analagously, we define the \emph{Moyal bracket}
\begin{equation}
	\left\{f,g\right\}  _{mb}=\frac{1}{i\hbar}(f\star g-g\star f)={\text{symb}}\,\left(  \frac{1}{i\hbar}\left[  \widehat{f},\widehat{g}\right]\right)
	\label{eq:moyal_bracket}
\end{equation}
as the symbol corresponding to the conmutator in $\widehat{{\mathcal{A}}}$.

In the limit $\hbar\rightarrow0$,
\begin{align}
	& (f\star g)(\phi)=f(\phi)g(\phi)+{\mathcal{O}}(\hbar) \\
	& \left\{  f,g\right\}  _{mb}=\left\{  f,g\right\}  _{pb}+{\mathcal{O}}(\hbar^{2})
	\label{eq:hzero_limit}
\end{align}
so the star product becomes the ordinary product and the Moyal bracket becomes the Poisson bracket. Moreover, it can be proved that, if $\widehat{f}$ commutes with $\widehat{g}$, Eqs. (\ref{eq:symb_product}) and (\ref{eq:hzero_limit}) give
\begin{equation}
	(f\star g)(\phi)=f(\phi)g(\phi)+{\mathcal{O}}(\hbar^{2}).
	\label{eq:symb_product2}
\end{equation}

Finally, we must define a unique inverse $symb^{-1}$ to complete the one-to-one mapping. We choose the symmetrical or \emph{Weyl ordering prescription}
\begin{align}
	&  {\text{symb}}\,^{-1}:{\mathcal{A}}_{q}\rightarrow\widehat{{\mathcal{A}}} \\
	&  {\text{symb}}\,^{-1}(qp)=\frac{1}{2}(\widehat{q}\widehat{p}+\widehat{p}\widehat{q}).
	\label{eq:symb_inverse}
\end{align}
This isomorphism established between the `phase-space' and `ordinary' quantum structures is the \emph{Weyl-Wigner-Moyal symbol}.

The WWM symbol for any $\widehat{\rho}\in\widehat{{\mathcal{A}}}^{\prime}$, $\widehat{{\mathcal{A}}}^{\prime}$ being the dual of $\widehat{{\mathcal{A}}}$, is defined as
\begin{equation}
	\rho(\phi)={\text{symb}}\,\widehat{\rho}=(2\pi\hbar)^{-N}{\text{symb}}\,_{({\text{for operators}})}\widehat{\rho}.
	\label{eq:symbol_dual}
\end{equation}
From this definition, we have
\begin{equation}
	\left\langle \widehat{O}\right\rangle =\text{tr}({\widehat{\rho}\widehat{O}})=\int d\phi^{2N}\rho(\phi)O(\phi).
	\label{eq:symbol_transition}
\end{equation}

In particular, consider the description of a quantum system in the EID formalism. The decomposition of the whole system ${\mathcal{U}}={\mathcal{S}}\otimes{\mathcal{E}}$ suggests a distinction between the proper system coordinates and the environmental ones. If $(x_{1},x_{2},...x_{N_{S}})$ and $(\chi_{1},\chi_{2},...\chi_{N_{E}})$ are the position coordinates of the relevant system and the environment, respectively, then the position coordinates of the universe are $(X_{\alpha})$, where $\alpha=1,2,...N$ and $N={N_{S}}+{N_{E}} $. Analogously, the momentum coordinates are $(p_{1},p_{2},...p_{N_{S}},\pi_{1},\pi_{2},...\pi_{N_{E}})=(P_{\alpha})$. Then the WWM transformation of a EID operator $\widehat{O}_{R}=\widehat{O}_{\mathcal{S}}\otimes\widehat{I}_{\mathcal{E}}$ reads
\begin{equation}
	 {\text{symb}}\,\widehat{O}_{R}=O_{R}(X_{\alpha},P_{\beta})=\int\langle{X_{\alpha}+\Delta_{\alpha}}|{\widehat{O}_{R}}|{X_{\alpha}-\Delta_{\alpha}}\rangle\operatorname{exp}{\left( i\frac{\Delta_{\alpha}P_{\beta}}{2\hbar}\right)}d\Delta^{N}
	\label{eq:OR_symbol}
\end{equation}
namely
\begin{align}
	& {\text{symb}}\,\widehat{O}_{R}=O_{R}(X_{\alpha},P_{\beta})=\int\langle{x_{i}+\Delta_{i}}|{\widehat{O}_{\mathcal{S}}}|{x_{i}-\Delta_{i}}\rangle\operatorname{exp}{\left( i\frac{\Delta_{i}p_{i}}{2\hbar}\right)}d\Delta^{N_{S}}\times\nonumber\\
	& \int\langle{\chi_{i}+\Delta_{i}}|{\widehat{I}_{\mathcal{E}}}|{\chi_{i}-\Delta_{i}}\rangle\operatorname{exp}{\left(i\frac{\Delta_{i}\pi_{i}}{2\hbar}\right) }d\Delta^{N_{E}}
	\label{eq:OR_symbol2}
\end{align}
But $\langle{\chi_{i}+\Delta_{i}}|{\widehat{I}_{\mathcal{E}}}|{\chi_{i}-\Delta_{i}}\rangle=\delta(2\Delta_{i})$ so
\begin{equation}
	\int\langle{\chi_{i}+\Delta_{i}}|{\widehat{I}_{\mathcal{E}}}|{\chi_{i}-\Delta_{i}}\rangle\operatorname{exp}{\left(i\frac{\Delta_{i}\pi_{i}}{2\hbar}\right) }d\Delta^{N_{E}}\sim1
\end{equation}
and thus with and adequate normalization the last integral can be considered equal to one. As a consequence, the environment formally disappears from the integral. The transformation yields
\begin{equation}
	{\text{symb}}\,\widehat{O}_{R}=O_{R}(X_{\alpha},P_{\beta})={\text{symb}}\,\widehat{O}_{\mathcal{S}}=O_{\mathcal{S}}(x_{i},p_{j}).
	\label{eq:symb_rel_operators}
\end{equation}

As $\widehat{\rho}_{\mathcal{S}}$ is a functional over the space of the $\widehat{O}_{\mathcal{S}}$ we have
\begin{equation}
	{\text{symb}}\,\widehat{\rho}_{\mathcal{S}}=\rho_{\mathcal{S}}(x_{i},p_{j}).
	\label{eq:symb_state}
\end{equation}
Then for Eq. (\ref{eq:symbol_transition}) we obtain
\begin{equation}
	\left\langle \widehat{O}_{R}\right\rangle _{\widehat{\rho}}=\text{tr}({\widehat{\rho}\widehat{O}_{R}})=\text{tr}({\widehat{\rho}_{\mathcal{S}}\widehat{O}_{\mathcal{S}}})=({{\text{symb}}\,\widehat{\rho}_{\mathcal{S}}}|{{\text{symb}}\,\widehat{O}_{\mathcal{S}}})=\int\rho_{\mathcal{S}}O_{\mathcal{S}}\,dx^{N_{S}}dp^{N_{S}}.
	\label{eq:symb_partial}
\end{equation}

\subsection{Fundamental graininess}
\label{sec:graininess}

The quantum uncertainty relations are obviously inconsistent with the classical idealization of a `representative point' in the phase-space where both position and momentum coordinates are determined with perfect precision. Indeed, in the quantum case we can define `small boxes' of volume $(\sigma_{x}\sigma_{p})^{d}$, proportional to the product of the variances of the positions $\sigma_{x}$ and momentums $\sigma_{p}$, respectively, and where $2d$ is the dimension of the phase space. This `fundamental graininess' characterized by the finite size boxes would be the analogue to the classical notion of point.

Once the to-become-classical properties are found, i.e. that the preferred states and its symbols are determined, we must study the evolution of its mean values over the fundamental boxes. For example, when $d=1$, If $f(x,p)\in\mathcal{A}_{q}$ is the symbol of some preferred state $\widehat{f}\in\widehat{\mathcal{A}}$, then we can compute its mean value in a box `$\sigma_{x}\sigma_{p}$' when the system is in the state $\widehat{\rho}_{\mathcal{S}}$ as
\begin{equation}
	\overline{f}(x,p,t)=\frac{1}{\sigma_{x}\sigma_{p}}\int\limits_{\sigma_{x}\sigma_{p}}f(x,p)\rho_{\mathcal{S}}(x,p,t)dxdp
	\label{eq:symb_meanvalue}
\end{equation}
with ${\text{symb}}\,\rho_{\mathcal{S}}(x,p,t)=$ symb $\widehat{\rho}_{\mathcal{S}}$.

\section{The classical limit}
\label{sec:classical_limit}

In previous papers, where we have introduced a general definition for the MPB $\left\vert i_{PS}(t)\right\rangle $ based in the pole theory, we show that, even when $\widehat{\rho}_{\mathcal{S}}$ evolves in a non-unitary way, the basis $\left\{  \left\vert i_{PS}(t)\right\rangle \right\}$ is always orthonormal so it must evolve unitarily as
\begin{equation}
	\left\vert i_{PS}(t)\right\rangle =\operatorname{exp}{\left(  -\frac{i}{\hbar}\int_{0}^{t}\widehat{\aleph}(t^{\prime})dt^{\prime}\right)  }\left\vert i_{PS}(0)\right\rangle
	\label{eq:basis_evol}
\end{equation}
where $\widehat{\aleph}(t)=\widehat{\aleph}(t)^{\dag}$ would be a time dependent effective Hamiltonian for the system ${\mathcal{S}}$. But, as previously remarked, $\widehat{\rho}_{PS}(t)$ does not evolve unitarily even if the projectors
\begin{equation}
	\widehat{\Pi}_{i}(t)=\left\vert {i_{PS}(t)}\right\rangle \left\langle{i_{PS}(t)}\right\vert =\text{e}^{-\frac{i}{\hbar}\int_{0}^{t}\widehat{\aleph}(t^{\prime})dt^{\prime}}\left\vert {i_{PS}(0)}\right\rangle \left\langle{i_{PS}(0)}\right\vert\text{e}^{\frac{i}{\hbar}\int_{0}^{t}\widehat{\aleph}(t^{\prime})dt^{\prime}}
	\label{eq:proj_evol}
\end{equation}
do. Then only these projectors $\widehat{\Pi}_{i}$ and their linear combinations (with constant coefficients) evolves unitarily. Precisely,
\begin{equation}
    \begin{split}
    \frac{d}{dt}\widehat{\Pi }_{i}(t)=-\frac{i}{\hbar }\text{e}^{-\frac{i}{\hbar}\int_{0}^{t}\widehat{\aleph }(t^{\prime })dt^{\prime }}\widehat{\aleph}(t)\left\vert {i_{PS}(0)}\right\rangle \left\langle {i_{PS}(0)}\right\vert\text{e}^{\frac{i}{\hbar }\int_{0}^{t}\widehat{\aleph}(t^{\prime})dt^{\prime }}\\ + \frac{i}{\hbar }\text{e}^{-\frac{i}{\hbar }\int_{0}^{t}\widehat{\aleph}(t^{\prime})dt^{\prime}}\left\vert {i_{PS}(0)}\right\rangle \left\langle{i_{PS}(0)}\right\vert \widehat{\aleph}(t)\text{e}^{\frac{i}{\hbar}\int_{0}^{t}\widehat{\aleph }(t^{\prime })dt^{\prime }}=-\frac{i}{\hbar }[\widehat{\aleph }(t),\widehat{\Pi}_{i}(t)]\text{.}
    \end{split}
    \label{eq:mpb_evoll}
\end{equation}

So the projectors $\widehat{\Pi}_{i}\in\widehat{{\mathcal{A}}}$ evolve as states. Both $\{\widehat{\Pi}_{i}\}$ and $\widehat{\aleph}(t)$ are operators in $\widehat{\mathcal{A}}$, thus we can make theirs WWM transformations, in ${\mathcal{S}}$, as (see Eq. (\ref{eq:symb_state}))
\begin{align}
	&  \aleph(x,p,t)={\text{symb}}\,\widehat{\aleph}(t)\text{ ,} \\
	&  \Pi_{i}(x,p,t)={\text{symb}}\,\widehat{\Pi}_{i}(t)\text{ .}
	\label{eq:mpb_symbols}
\end{align}
Moreover, we can use the result of Eq. (\ref{eq:symb_product2}) for the WWM transformation of a product of commuting operators to deduce that
\begin{equation}
	 {\text{symb}}\,\widehat{\Pi}_{i}={\text{symb}}\,(\widehat{\Pi}_{i}\widehat{\Pi}_{i})=\Pi_{i}^{2}(x,p)+{\mathcal{O}}(\hbar^{2})=\Pi_{i}(x,p)+{\mathcal{O}}(\hbar^{2})\text{ .}
	\label{proj_prod}
\end{equation}
So, in the limit $\hbar^{2}\rightarrow0$ we have
\begin{equation}
	\Pi_{i}(x,p)\left[  \Pi_{i}(x,p)-1\right]  =0\text{ .}\label{eq:classic_proj}
\end{equation}
Thus, $\Pi_{k}(x,p)$ is a characteristic function in ${\mathcal{A}}_{q}$ that defines a certain domain ${\mathcal{D}}_{k}$ such that
\begin{equation}
	\Pi_{k}(x,p)=
	\begin{cases}
		1 & {\text{if }}(x,p)\in{\mathcal{D}}_{k};\\
		0 & {\text{if }}(x,p)\notin{\mathcal{D}}_{k}.
	\end{cases}
	\label{eq:classic_proj2}
\end{equation}

We will assume, without lost of generality, that the domains ${\mathcal{D}}_{k}$ are \emph{connected}. We will later give the motivations of this assumption. Also, the product of two different projectors vanish
\begin{equation}
	\widehat{\Pi}_{i}(t)\widehat{\Pi}_{j}(t)=\left\vert {i_{PS}(t)}\right\rangle\left\langle {i_{PS}(t)}|{j_{PS}(t)}\right\rangle\left\langle {j_{PS}(t)}\right\vert =0\hspace{0.5cm}{\text{if }}i\neq j\text{ ,}
	\label{eq:proj_prod}
\end{equation}
then when $\hbar\rightarrow0$ we have
\begin{equation}
	\Pi_{i}(x,p,t)\Pi_{j}(x,p,t)=0\hspace{0.5cm}{\text{if }}i\neq j
	\label{eq:proj_prod_class}
\end{equation}
so the corresponding domains of the projectors are disjoint, i.e.
\begin{equation}
	{\mathcal{D}}_{i}\cap{\mathcal{D}}_{j}=\emptyset\hspace{0.5cm}{\text{if }}i\neq j\text{ .}
	\label{eq:disjoint_domains}
\end{equation}

Now we can write the equation (\ref{eq:preferred_state}) as
\begin{equation}
	\widehat{\rho}_{PS}(t)=\sum_{k}p_{k}(t)\widehat{\Pi}_{k}(t)
	\label{eq:ps_proj}
\end{equation}
and its corresponding `classical version'
\begin{equation}
	\rho_{PS}(x,y,t)=\sum_{k}p_{k}(t)\Pi_{k}(x,y,t)
	\label{eq:ps_proj_class}
\end{equation}
where each $\Pi_{k}(x,y,t)$ corresponds to a `moving state' (the WWM transform of the MPB projectors) with domain ${\mathcal{D}}_{k}$. The spectral decompositions (Eqs(\ref{eq:ps_proj,eq:ps_proj_class})) can be discrete or continuous. In the classical usual case with discrete spectrum\footnote{The continuous case needs for a mathematically rigorous formulation in a completely different formalism.}, and if we have normalized the volume of phase space to be one, we can decompose the phase space of the system in a finite sum of domains of finite measure, i.e. (see Eq.(\ref{eq:classic_proj2}))
\begin{equation}
	{\text{Vol}}({\mathcal{D}}_{i})\geq0\hspace{0.2cm}\forall i,\hspace{0.5cm} \sum_{i}{\text{Vol}}({\mathcal{D}}_{i})\leq1
	\label{phasespace_decomp}
\end{equation}
where `Vol' refers the corresponding volume computed with the metric of the phase space of ${\mathcal{S}}$.

In order to define the basis $\{\left\vert {i_{PS}(t)}\right\rangle \}$ we need $N_{\mathcal{S}}$ indices $i$ (as in the H atom in 3 dimensions we need three indices $[\omega,l,m]$ and three observables $[\widehat{H},\widehat{L}^{2},\widehat{L}_{z}]$). Thus, this is the index of projector $\widehat{\Pi}_{i}$. E.g., from ref. (paper sist cerrados) we know that, in the closed system case, the role of the projectors is played by the constants of the motion ($[\widehat{H},\widehat{L}^{2},\widehat{L}_{z}]$ in the H atom) that are defined by
\begin{equation}
	\lbrack\widehat{H},\widehat{C}_{i}]=0\text{ .}
	\label{eq:constants_motion}
\end{equation}
Instead, the $\widehat{\Pi}_{i}$ of open systems are not strictly constant and they satisfy Eq. (\ref{eq:proj_evol2})
\begin{equation}
	\frac{d}{dt}\widehat{\Pi}_{i}(t)=-\frac{i}{\hbar}[\widehat{\aleph}(t),\widehat{\Pi}_{i}(t)]
	\label{eq:mpb_commutator}
\end{equation}
and even if the commutator is not zero, since the $\widehat{\Pi}_{i}$ are chosen by the criteria listed in section (\ref{sec:mpb_criteria}), they are those that minimized this commutator. This is the best possible choice for the role of `action variables', since the $\{\widehat{\Pi}_{i}\}$ turn out to be the `most constant' operators involved in the non-unitary evolution of the state of the relevant system. The operators that play the role of the `angular variables' $\widehat{\Phi}_{i}$ are the corresponding conjugated variables
\begin{equation}
	\lbrack\widehat{\Phi}_{i}(t),\widehat{\Pi}_{j}(t)]=i\hbar\delta_{ij}\widehat{I}\hspace{0.5cm}\forall\,i,j
	\label{eq:angular-variables}
\end{equation}
where $\widehat{I}$ is the identity operator in ${\mathcal{S}}$. This last equation is the definition of the operators $\widehat{\Phi}_{i}$. The WWM transformation of this equation gives
\begin{equation}
	\{\Phi_{i}(x,p,t),\Pi_{j}(x,p,t)\}_{pb}=\delta_{ij}+{\mathcal{O}}(\hbar^{2})
	\label{eq:wwm_angular_var}
\end{equation}
with $\Phi_{i}(x,p,t)={\text{symb}}\,\widehat{\Phi}_{i}(t)$. Therefore, these variables form a set $\{\Phi_{i}(x,p,t),\Pi_{j}(x,p,t)\}$ of `action-angle' variables that become canonical conjugated in the limit $\hbar\rightarrow0$.

Now, the dynamics of the $\{\widehat{\Phi}_{i}(t),\widehat{\Pi}_{i}(t)\}$ operators satisfy
\begin{equation}
	\begin{cases}
	\frac{d}{dt}\widehat{\Phi}_{i}(t) & =-\frac{i}{\hbar}[\widehat{\aleph}(t),\widehat{\Phi}(t)]\hspace{0.5cm}\forall i,\\
	\frac{d}{dt}\widehat{\Pi}_{i}(t) & =-\frac{i}{\hbar}[\widehat{\aleph}(t),\widehat{\Pi}(t)]\hspace{0.5cm}\forall i
	\end{cases}
	\label{eq:action_angle_dynamics}
\end{equation}
and making the WWM transformation of these equations we obtain the dynamics of the $\{\Phi_{i}(x,p,t),\Pi_{i}(x,p,t)\}$ variables
\begin{equation}
	\begin{cases}
	\frac{d}{dt}\Phi_{i}(x,p,t) & =\{\aleph(t),\Phi(t)\}_{pb}+{\mathcal{O}}(\hbar^{2})\hspace{0.5cm}\forall i,\\
	\frac{d}{dt}\Pi_{i}(x,p,t) & =\{\aleph(t),\Pi(t)\}_{pb}+{\mathcal{O}}(\hbar^{2})\hspace{0.5cm}\forall i.
	\end{cases}
	\label{eq:phase_space_dynamics}
\end{equation}
Then, as $\hbar\rightarrow0$, since the evolution of the $\widehat{\Pi}_{i}$ is adiabatic while the one of $\widehat{\Phi}_{i}$ is not adiabatic, we have from equations (\ref{eq:action_angle_dynamics}) and (\ref{eq:phase_space_dynamics})
\begin{equation}
	\begin{cases}
	\frac{d}{dt}\Phi_{i}(x,p,t)\approx0\hspace{0.5cm}\forall i,\\
	\frac{d}{dt}\Pi_{i}(x,p,t)\neq0\hspace{0.5cm}\forall i.
	\end{cases}
	\label{eq:phase_space_limitdynamics}
\end{equation}
So, the $\Pi_{i}$'s in ${\mathcal{A}}_{q}$ are also approximately constants, while the $\Phi_{i}$'s are variables.

To obtain the classical trajectories we must solve the system of differential equations (\ref{eq:phase_space_limitdynamics}) with the corresponding initial conditions $\{\Phi_{i}(x,p,0),\Pi_{i}(x,p,0)\}$. In fact, if we consider the initial conditions as arbitrary data we have
\begin{equation}
	\begin{cases}
	\Phi_{i}(x,p,t)=\mathcal{F}[\Phi_{i}(x,p,0),\Pi_{i}(x,p,0);t]=\text{`variable',}\\
	\Pi_{i}(x,p,t)=\mathcal{G}[\Phi_{i}(x,p,0),\Pi_{i}(x,p,0);t]\approx\text{`constant'.}
	\end{cases}
	\label{eq:traj_initial}
\end{equation}
Finally, we must proceed as explained in \ref{sec:graininess}. We choose the initial conditions to be different from zero only in a box and we obtain (see Eq.(\ref{eq:symb_meanvalue}))
\begin{equation}
\begin{cases}
	 \overline{\Phi}_{i}(t)=\frac{1}{\sigma_{x}\sigma_{p}}\int\limits_{\sigma_{x}\sigma_{p}}\mathcal{F}[\Phi_{i}(x,p,0),\Pi_{i}(x,p,0);t]\rho_{\mathcal{S}}(x,p,t)dxdp=\text{`variable',}\\
	 \overline{\Pi}_{i}(t)=\frac{1}{\sigma_{x}\sigma_{p}}\int\limits_{\sigma_{x}\sigma_{p}}\mathcal{G}[\Phi_{i}(x,p,0),\Pi_{i}(x,p,0);t]\rho_{\mathcal{S}}(x,p,t)dxdp\approx\text{`constant'.}
	\end{cases}
	\label{eq:traj}
\end{equation}
It is worthwhile to note that the mean value over the selected box will be different from zero only when the box is \emph{in} the domain $\mathcal{D}_i$ of the corresponding $\Pi_i$. These `action-angle' variables will give the simplest equations for the trajectories. We could recover the description in the position-momentum space by a canonical transformation that links both descriptions: $\{(\Phi_{i},\Pi_{i});\aleph_{(\Phi,\Pi)}\}\leftrightarrow\{(x_{i},p_{i});\aleph_{(x,p)}^{\prime}\}$.

\begin{figure}
	\begin{center}
	    \includegraphics[width=.7\columnwidth]{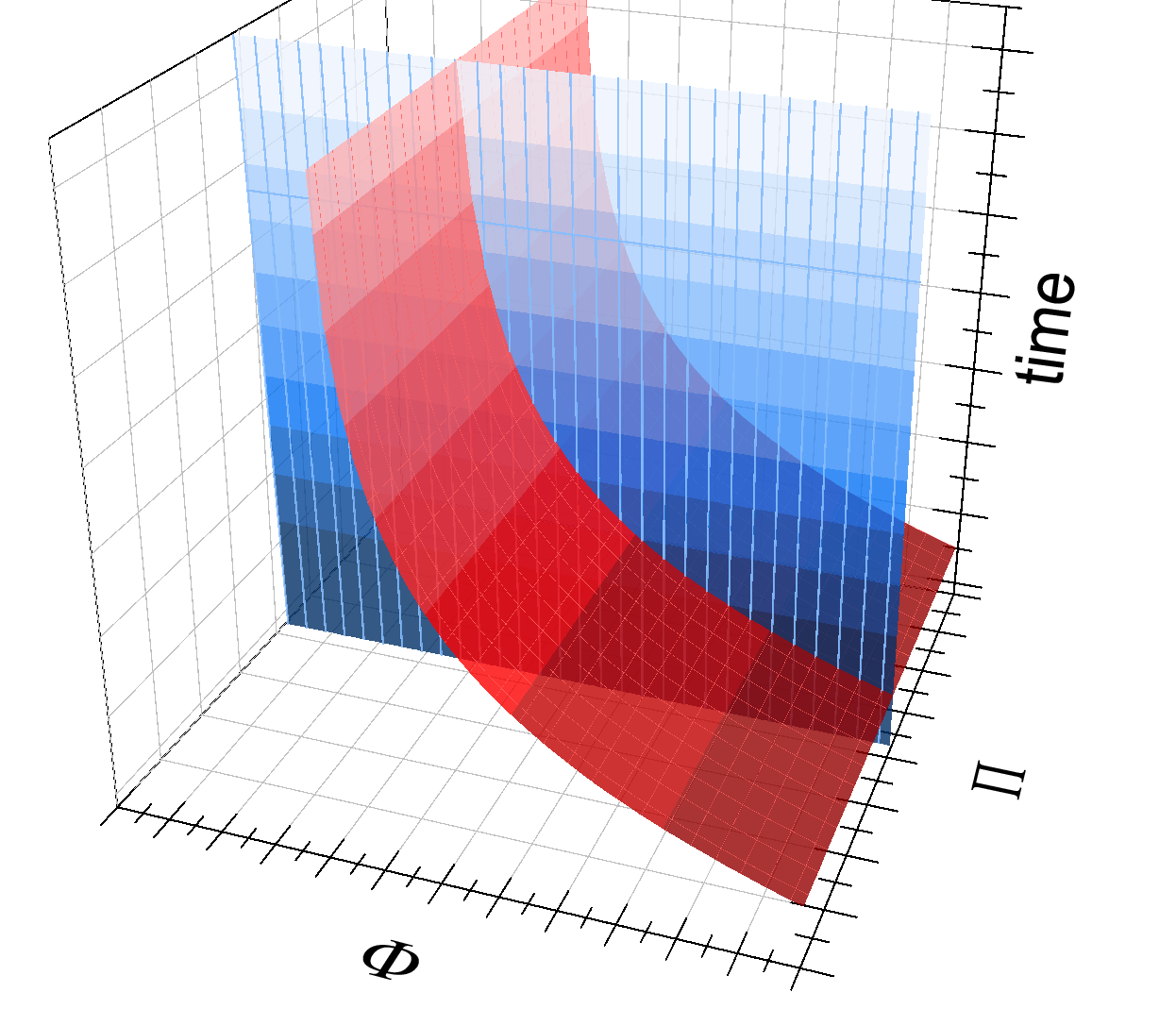}\\
	\caption{The intersection of the surfaces defined by the transformation of the MPB and its conjugated set into the phase-space gives, in certain limit, the emergence of the classical trajectories. Given the non-unitary character of the evolution of the proper system, the trajectory could reach an equilibrium limit as it is showed in this case.}
    \label{fig:evol}
	\end{center}
\end{figure}

The simplest case would be the one of an open system in one dimension, so we have the phase-space coordinates $(\Pi(t),\Phi(t))$. Lets call $\varphi(t)$ the function that results from the integrals for $\overline{\Phi}$ in Eq.(\ref{eq:traj}). In the $(\Pi,\Phi,t)$ space both the relations $\overline{\Pi}(t)\approx$`constant' and $\overline{\Phi}=\varphi(t)$ defines two dimensional surfaces $S_{\Pi}=\{(\Pi,\Phi)/\overline{\Pi}(t)\approx$`constant' and $S_{\Phi}=\left\{  (\Pi,\Phi)/\overline{\Phi}(t)\approx\varphi(t)\right\}$, respectively. The intersection $S_{\Pi}\cap S_{\Phi}$ gives the trajectories in the phase space. In Fig.\ref{fig:evol} we represent a case where the trajectory has an equilibrium limit, that is a possibility since the evolution is non-unitary.

Reasonably, we can consider that the $\widehat{\Pi}_i(t)$ actualize (`reduce' or `collapse') and then they begin to belong to the classical realm as the classical functions $\Pi_i(x,p,t)$. Indeed, as we are in the limit $\hbar\rightarrow 0$ where the indetermination principle gives $\sigma_x\sigma_p\sim 0$, the classical values are simply given by $\overline{\Pi}_i(t)$ (Eq.(\ref{eq:traj})). But in this limit also the $\widehat{\Phi}_i(t)$ actualize and they begin to belong to the classical realm as the functions $\Phi_i(x,p,t)$, with classical values $\overline{\Phi}_i(t)$.

\section{Summary and conclusions}
\label{sec:conclusions}

Essentially, using the EID formalism combined with the definition of the Moving Preferred Basis we have shown that it is possible to recover the classical trajectories that satisfy the correspondence principle in the limit $\hbar\rightarrow 0$. Moreover, we have shown that this trajectories, that could be represented in the phase-space of the relevant (open) system $\mathcal{S}$, can reach equilibrium points.

May be this result could seem more or less trivial, but it is not so because it only refers to a particular basis (the MPB) and to its associated projectors ($\widehat{\Pi}_i$), while the state of the system, when $t>t_D$, is $\widehat{\rho}_{\mathcal{S}}(t)=\sum_{k}p_{k}(t)\widehat{\Pi}_{k}(t)$, where the $p_k(t)$ are real functions of the time such that $0\leq p_k(t) \leq 1$. So, if at some time $t_0$, it may be that $p_k(t_0)=0$, in this case the actualization is impossible, or it may be that $p_k(t_0)=1$ and in this case the actualization necessarily take place at $t_0$ and the state becomes $|k_{PS}(t_0)\rangle$. These results, and probably others of this kind that may be found, are not trivial at all.

Then, we hope to have opened a line of research which deserves some attention and, of course, we have to continue our research to base our results on many other models.

\section*{Aknowledgments}

We are very grateful to Roberto Laura, Olimpia Lombardi, Roland Omn\`es and Maximilian Schlosshauer for many
comments and criticisms. This research was partially supported by grants of the University of Buenos Aires, the
CONICET and the FONCYT of Argentina.

\section{References}
	\bibliographystyle{ieeetr}
	\bibliography{decoherence}

\end{document}